# An Approach Towards Intelligent Accident Detection, Location Tracking and Notification System

Supriya Sarker,[1,*] Md. Sajedur Rahman,[2] and Mohammad Nazmus Sakib[3]

[1]Department of Computer Science & Engineering, Chittagong University of Engineering & Technology
Chittagong, Bangladesh

[2]Department of Electrical & Electronic Engineering, [3]Department of Computer Science & Engineering

[2,3]World University of Bangladesh

Dhaka, Bangladesh

[*]sarkersupriya7@gmail.com

*Abstract*—**Advancement in transportation system has boosted speed of our lives. Meantime, road traffic accident is a major global health issue resulting huge loss of lives, properties and valuable time. It is considered as one of the reasons of highest rate of death nowadays. Accident creates catastrophic situation for victims, especially accident occurs in highways imposes great adverse impact on large numbers of victims. In this paper, we develop an intelligent accident detection, location tracking and notification system that detects an accident immediately when it takes place. Global Positioning System (GPS) device finds the exact location of accident. Global System for Mobile (GSM) module sends a notification message including the link of location in the google map to the nearest police control room and hospital so that they can visit the link, find out the shortest route of the accident spot and take initiatives to speed up the rescue process.**

## I. INTRODUCTION

The evolution of the transportation system has made possible the rapid change in the civilization of human history. Transportation has great importance in our daily life and its development has made our communication much easier. But it can be also cause of loss of lives and properties. According to Association for Safe International Road Travel [1] approximately 1.25 million people die in road accident every year while 3,287 deaths every day. Besides, 20-50 million people are injured and disabled. Road traffic accident is ranked as 9[th] leading cause of loss of lives. Annually USD $518 billion has been damaged because of road traffic accident which costs from 1-2% of annual GDP of individual countries. The most obvious reason of a person's death during accidents is the unavailability of the first aid provision due to the delay in accessing the exact information of the accident location. The trivial reasons of unavailability of prompt on-site medical assistance are late reporting of accident, incorrect geographic location [2], and unreliable mobile application due to malfunctioning of smartphone [3]. The typical scenario is eyewitnesses do not notify hospital and police while an accident takes place. Moreover, accidents occur on highways, vehicles passing the accident location generally ignore their responsibility to notify hospital and police [4]. Hence, this problem needs immediate attention and intelligent accident detection and location tracking system should be developed which do not required human intervention.

The paper begins by discussing the significance of a lightweight, standalone, intelligent accident detection system in the current scenario. A brief description of the existing approaches is presented in Section II. The methodology of the proposed system including system architecture is described in Section III. Hardware set up is described in Section IV. The working flow diagram is shown in Section V. The experimental results along with the device prototype are presented in Section VI. Finally, Section VII concludes the paper with future direction.

## II. EXISTING APPROACH

From the literature review, it is clear that enormous research has been done to detect an accident and notify the emergency medical services. In [5] the speed of the vehicle has monitored with the help of GPS receiver and detected accident by comparing the previous and current speed of vehicles. This system sent the location of accident to an Alert Service Center. However, the distance between accident location and Alert Service Center might affect the time required to rescue or to send help. In [4] the authors proposed a system that notified emergency contact, police or ambulance services while an accident has detected. The system programmed a smartphone for decision support based on Mamdani Fuzzy Logic to detect accident and stored records in the data center for future use. The data regarding accidents were collected using Accelerometer, Gyroscope and four force sensors attached at each side of the vehicle. However, the system hardware with excessive sensors was not compact enough. Though the system sent a text message to the emergency contact or public safety, the response time depends on the distance between the location of the accident and the public safety service center. In [2] the authors employed a shock sensor to detect accident and sent geographic location, additional information of passengers such as full name, blood type, phone number, email, medical history, date of birth, and reference phone number to the headquarters of the Public Safety Organizations. The main limitation of the system is all the passengers and vehicles need to be pre-registered with their information which is not practical for the current scenario of Bangladesh. Besides, the success of the system relied upon the distance between the headquarters of the

Public Safety Organizations and accident location. In [6] a map matched system has been proposed to locate the vehicle using data collected by GPS. The system detected accident by comparing the previous and current speed of the vehicle. While an accident detected it sent the location to an Alert Service Center. A crash detection using acceleration data and On-board Diagnostics system has been implemented to receive the status of the vehicle and notify emergency medical services in [3]. But the OBD system is not available in every vehicle in Bangladesh. An intelligent accident detection system using the tilt of the vehicle and abnormal heartbeat rate of the passengers has been proposed in [7]. A smartphone received the data from an accelerometer and heartbeat sensor detected accident if the threshold value has exceeded and sent SMS to the phone numbers stored in that smartphone. Along with this, it searched for the contact number of nearby Emergency Medical Assistance. However, sending SMS to all the contact numbers stored in the smartphone would be redundant and increased system overhead. Possibilities of the glitches in the smartphone due to battery issues may make the system unreliable and activation of the smartphone application every time, when the driver starts the journey makes it absurd [2]. Message Queuing Telemetry Transport (MQTT) was used to send an alert through email while a vehicle accident has detected and accident data received by accelerometer and ultrasonic sensors stored in the Losant IoT platform in [8]. However, this system could not detect the exact location of the accident. Hence, rapid rescue and sending help is quite difficult. Email communication is comparatively slower than a text message on a smartphone. RFID based real-time accident detection algorithm has been employed in [9] which collected information such as speed, vehicle plate number and the number of passengers of the vehicle though it is not necessary to collect the number of passengers in the vehicle. In this system, four crash sensors were mounted in the four corners of the vehicle which made the system less compact.

Hence, the major contribution made in this research is the development of a lightweight, compact design intelligent vehicle accident detection, location tracking and notification system which send a location from Google map to nearby hospital and police station. This Google map link shows the location of the accident and helps to find the optimum route to reach in the location as well.

### III. PROPOSED METHODOLOGY

This section discusses the architecture and functionality of the proposed system. Fig. 1. demonstrates the architecture of the proposed system. The proposed has divided into three phases which are: 1) Accident Detection, 2) Location Tracking, and 3) Notification Sending.

*1) Accident Detection*: An accelerometer sensor senses the accident when the vehicle is fallen down in any direction of X, Y, Z-axis. Initially, the angle of the vehicle is zero degree and it could be increased up to 360 degree towards any axis. If the angle of the vehicle rises in any direction exceeds our threshold value the accelerometer considers the situation as an accident. The threshold value in the X and Y axis are 310 and 340, respectively. If an accident is detected, then the sensor will send the signal to the microcontroller. We have also used two ultrasonic sensors in front and back of the vehicle. The ultrasonic sensor is always turn on when an object reaches within 5 cm of the vehicle which sometimes creates a false prediction of collision.

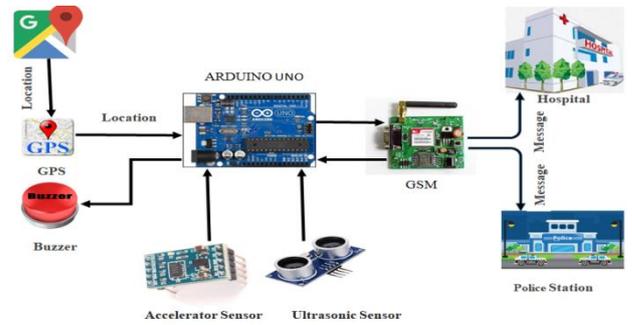

Fig. 1 Architecture of intelligent accident detection, location tracking and notification.

*2) Location Tracking*: The GPS sensor can detect the current location of the vehicle. In our proposed system we use the GPS device to find the exact accident location. When the microcontroller receives any signal of an accident it requests for the current location of the accident spot to the GPS. The GPS sends the location of the accident spot to the microcontroller.

*3) Notification Sending:* With accident location link GSM sends a text message to the hospital and police control room. The hospital and police control room will get a message along with the map link which will contain the exact latitude and longitude details of the location. At the same time, the nearest police station receives a notification message of accident with a link Google map. With the help of these details, the ambulance can take the shortest route to the accident location and reduce the time to save the lives of the victims.

### A. Hardware Set up

Circuit connection of our intelligent accident detection, location tracking, and notification system is very simple which is shown in fig. 2. The Tx pin of the GPS module is directly connected to the RX pin (RX0) of Arduino UNO. By using Software Serial Library, we have allowed serial communication on pin RX0 and TX0, and made them Rx and Tx respectively and left the Rx pin of the GPS Module open. Though generally pin 0 and pin 1 of Arduino UNO are used for serial communication, through the Software Serial library, we can allow serial communication on other digital pins of the Arduino as well. 5 Volt supply is used to operate the GPS Module. GSM module's Tx and Rx pins are directly connected to pin D8 and D7 of Arduino. For GSM interfacing, we also use a software serial library. GSM module is also powered by 5v supply. An optional LCD's data pins D4, D5, D6, and D7 are connected to pin number 12, 11, 6, and 5 of Arduino UNO. Command pin RS and EN of LCD are connected with pin 4 and pin 3 of Arduino UNO and the RW pin is directly connected with ground.

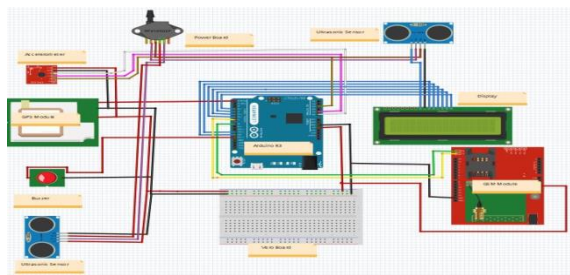

Fig. 2 Circuit Diagram of proposed intelligent accident detection, location tracking and notification system.

To adjust the contrast or brightness of LCD a potentiometer is used. An accelerometer sensor is added in the system for detecting an accident and its X, Y, and Z-axis ADC output pins are directly connected to Arduino ADC pin, A0, A1, and A2.

### B. Working Flow of the Proposed System

This section demonstrates the working flow of the proposed intelligent accident detection, location tracking and notification system. Fig. 3. shows the flowchart of the proposed system below:

## IV. RESULTS AND ANALYSIS

The device prototype and experimental results are discussed in this section. The performance and accuracy of the system is also being analysed as well.

### C. Device Prototype

Fig. 4 demonstrates the prototype of the proposed intelligent accident detection and alert system.

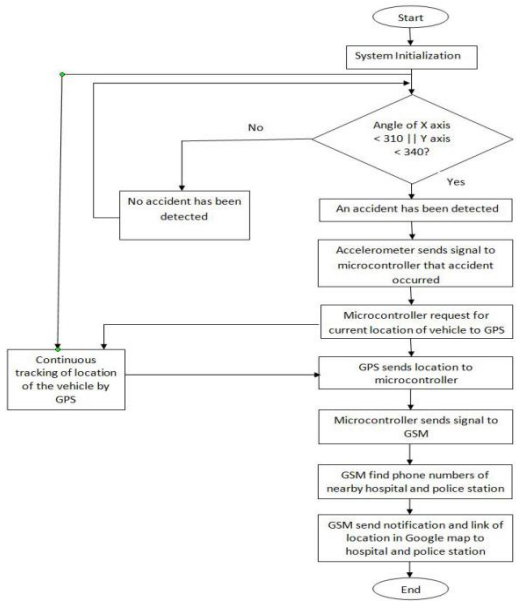

Fig. 3 Flowchart of intelligent accident detection, location tracking and notification system.

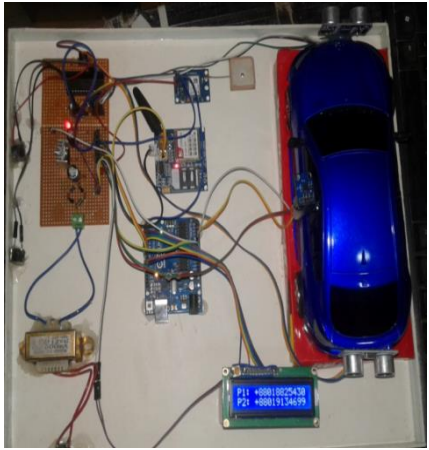

Fig. 4 Prototype of intelligent accident detection, location tracking and notification system.

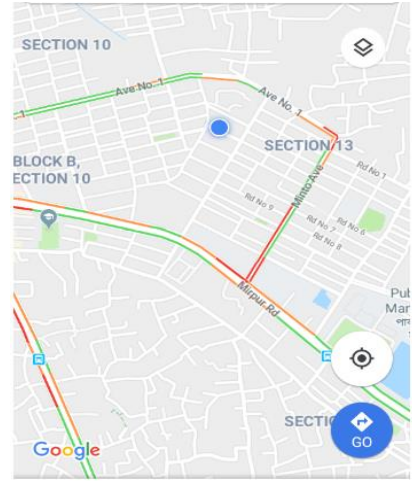

Fig. 5 Sending notification to nearby hospital and police station.

Two phone numbers are being displayed in the LED display which chosen by the proposed algorithm among the list of stored phone numbers of hospital and police station in the system beforehand. Fig. 5 is the pictorial representation of sending messages to the chosen numbers of nearby hospitals and police stations for rescue which is displayed in the LED display. Fig. 6 shows that the proposed algorithm tracks the accident location and fig. 7 shows that the system sends the link to the chosen phone number through message.

### D. Experimental Results

The intelligent accident detection and notification system has been tested in 20 different places in different cities of Bangladesh. We tested our system in sloping, uneven, curved highways for various degrees of angles. If the degree of angles in X or Y axis exceeded 310 and 340 degree, respectively for any case, the system detected it as an accident. If any vehicle reached within 5 cm, the system detected a possibility of crash. As the system has tested in highways there are lower possibilities of reaching any object, living being within 5 cm of the vehicle body. Table I listed the test results of the accident detection, exact location tracking and sending a notification message.

On the basis of the experimental data listed in Table 1, we evaluate system performance. The system performance is divided into three sections: (1) Accident detection accuracy, (2) Location tracking accuracy, (3) Notification sending accuracy.

*1) Accident Detection Accuracy*: Accident detection accuracy, $D_A$ is defined by (1)

$$D_A = \frac{TP_A}{TP_A + FN_A} \quad (1)$$

where, $TP_A$ and $FN_A$ denotes correct detection of accident and incorrect detection of accident, respectively.

*2) Location Tracking Accuracy:* Location tracking accuracy, $T_L$ is defined by (2)

$$T_L = \frac{TP_L}{TP_L + FN_L} \quad (2)$$

where, $TP_L$ and $FN_L$ denotes correct tracking of location and incorrect tracking of location, respectively.

TABLE I
EXPERIMENTAL RESULTS

| Place No. | Accident Detection Status | Exact Location Tracking Status | Notification Sending Status |
|---|---|---|---|
| 1 | Yes | Yes | Yes |
| 2 | Yes | Yes | Yes |
| 3 | Yes | Yes | Yes |
| 4 | Yes | Yes | Yes |
| 5 | Yes | Yes | Yes |
| 6 | No | Yes | Yes |
| 7 | Yes | Yes | Yes |
| 8 | Yes | Yes | Yes |
| 9 | Yes | No | Yes |
| 10 | Yes | Yes | Yes |
| 11 | Yes | Yes | Yes |
| 12 | Yes | Yes | Yes |
| 13 | Yes | Yes | Yes |
| 14 | Yes | Yes | Yes |
| 15 | Yes | No | Yes |
| 16 | Yes | Yes | Yes |
| 17 | Yes | Yes | Yes |
| 18 | Yes | Yes | Yes |
| 19 | Yes | Yes | Yes |
| 20 | Yes | Yes | Yes |

*3) Notification Sending Accuracy:* Notification Sending accuracy, $S_N$ is defined by (3)

$$S_N = \frac{TP_N}{TP_N + FN_N} \quad (3)$$

where, $TP_N$ and $FN_N$ denotes successful sending of notification and unsuccessful sending of notification in specified phone numbers, respectively.

Our proposed intelligent accident detection and notification system provide 95% accident detection accuracy, 90% exact location tracking accuracy and 100% successful notification sending accuracy in nearby hospital and police station.

## V. CONCLUSION

With the advent of science and technology in every walk of life, the importance of vehicle safety has received the highest priority. We design a low cost, portable vehicle accident detection system to resolve the problem of lack of an automated system for accident detection and location tracking. Consequently, the time for searching the location is reduced and therefore, the injured person can get help rapidly which will save many lives. The limitation of the system is the ultrasonic sensor which always turns on when an object reaches within 5 cm of the vehicle. Sometimes it creates a false prediction of collision. We assume that as the vehicle will be driven on the highway roads, there are lesser chances of reach of humans or objects other than vehicles around 5 cm of the front car frequently. Hence, the false positive rate will decrease. We have a future plan of an interconnecting camera to the controller module that will capture the images of the accident spot to facilitate easy accident detection and increase accountability towards maintaining traffic rules. The proposed system provides a feasible solution to traffic hazards and saves time and reduces the loss of lives.